\documentclass[runningheads]{llncs}

\usepackage[T1]{fontenc}
\usepackage{graphicx}
\usepackage{url}

\begin{document}

\title{Critsly: An Artefact-Aware AI Critique Teammate for Design Education and Project-Based Learning}

\titlerunning{Critsly: AI Critique Teammate}

\author{Nizam Kadir \and Juan David Salazar Rodriguez \and Sumaiyya Ali}

\authorrunning{N. Kadir et al.}

\institute{Singapore University of Technology and Design, Singapore\\
\email{\{nizam\_kadir, juandavid\_salazarrodriguez, sumaiyya\_ali\}@mymail.sutd.edu.sg}}

\maketitle

\begin{abstract}
Critique is central to design education and project-based learning, yet high-quality critique is often scarce, uneven, hard to document, and disconnected from evolving artefacts. We present Critsly, an artefact-aware AI critique workspace that turns AI from a detached feedback tool into a critique teammate in the learner's board context. Critsly combines a visual design canvas with structured AI-supported reflection, multi-perspective critique, action planning, optional peer/jury settings, and educator evidence traces. Unlike chatbot feedback tools that rely on isolated text prompts, Critsly grounds critique in a structured board state containing design intentions, board elements, annotations, links, and prior critique history. Reflecture, Critsly's guided reflection flow, works with Six Thinking Hats-inspired personas, board synthesis, generated action plans, exportable critique records, and educator evidence views in one workflow. The demo follows a learner from design intention to board-aware critique, persona-based evaluation, action-plan generation, and educator-facing evidence. Critsly contributes a working example of AI-supported critique that is more frequent, structured, inspectable, and actionable for learners and educators.
\keywords{Educational AI \and design education \and critique \and reflection \and project-based learning \and learning analytics}
\end{abstract}

\section*{Demo Link}

\url{https://www.youtube.com/watch?v=AbKcspm_hc8}

\section{Motivation}

Design education, architecture studio, UX education, and project-based learning rely heavily on critique~\cite{uluoglu2000}. Students improve by making their reasoning visible, receiving feedback, interpreting alternatives, and revising their artefacts. Yet critique is difficult to scale. It often happens episodically, depends on who is available, disappears after verbal discussion, and is hard for educators to inspect across a cohort.

Chatbot-based feedback tools can support reflection and writing, but they typically depend on prompt-supplied context rather than persistent board state and artefact structure~\cite{kasneci2023}. In design critique, the artefact matters: sketches, boards, sticky notes, design intentions, annotations, and revision history. Critsly therefore keeps critique attached to the student's work and uses board context to scaffold reflection rather than only deliver comments.

The AIED contribution is an interactive environment for critique as a human-AI learning process, aligned with the conference theme of moving from tools to teammates. In this setting, AI functions as a critique teammate that helps learners articulate intent, inspect evidence, consider multiple perspectives, and convert feedback into action, while educators retain judgment and receive evidence of critique participation and revision.

\section{System Presentation}

Critsly combines a visual canvas, an AI critique panel, structured reflection protocols, persona-based critique, and educator evidence surfaces. Its contribution lies in integrating board-grounded context, reflection scaffolds, perspective-taking personas, and reusable evidence within one critique workflow.

\textbf{Artefact-aware canvas.} Learners work on a board containing design artefacts such as sketches, images, sticky notes, text blocks, links, annotations, and connections. Critique is not separated from the work; the board is the context for conversation and analysis. When critique is requested, Critsly provides a structured board state---design-intention fields, board elements, annotations, links, and recent critique history---rather than only the learner's current prompt. This is board-grounded critique support, not automated assessment of every visual feature.

\textbf{Design intention anchoring.} Before critique, learners state the concept, goals, keywords, and readiness for feedback. This reduces vague critique by making the student's design intent explicit.

\textbf{Reflecture guided reflection.} Reflecture is Critsly's in-system guided reflection flow. It prompts description, evaluation, analysis, conclusion, and action planning to support critique practice.

\textbf{Six Thinking Hats-inspired personas.} Learners can request critique from different perspectives, such as evidence-focused, risk-focused, generative, emotional, optimistic, or process-oriented lenses~\cite{debono1985}. This helps students experience critique as a multi-perspective practice rather than a single authoritative judgment.

\textbf{Board analysis and action planning.} The system can synthesize board content, identify patterns or risks, and generate actionable next steps. Critique therefore becomes a revision workflow.

\textbf{Peer, jury, and educator evidence.} Critsly also supports optional anonymous peer or jury configurations. The core demo, however, centers on learner-AI critique and educator evidence indicators such as critique participation, persona selections, action-plan generation, and board revisions.

\section{Demonstrated Functionality}

The demo presents a representative design-studio workflow. A learner opens a project board, adds or reviews design artefacts, and records design intentions. The learner then requests AI critique using Reflecture. Critsly responds from the current board state rather than as a detached chatbot. The learner switches to a risk-focused persona, receives a different critique lens, and generates an action plan. The demo closes by showing export and educator evidence surfaces, illustrating how critique traces can support reflection, revision, and teaching decisions. Peer and jury critique are also supported, but the walkthrough focuses on the learner-AI critique cycle and the resulting evidence available to educators.

The system is implemented and has existing deployment evidence. The unified Critique Studio Canvas route integrates the canvas engine with the critique sidebar. The live system is available via \url{https://critsly.com}, where users can access the Critique Studio workflow. Project documentation records the unified interface, persona lenses, Reflecture, mobile capture, PII redaction, red-team evaluation, and multi-device support as completed. Production QA verified live board initialization, critique-panel operation, and canvas interaction on \path{/dashboard/studio-canvas/}.

\section{Originality, Educational Benefits, and Boundaries}

Critsly is organized around three design choices: artefact-aware critique on structured board context; structured critique via Reflecture and persona lenses; and evidence-producing workflows that preserve critique traces and action plans. Together, these features position the system as a learning environment for critique practice rather than merely a feedback generator.

The educational benefit is more frequent, inspectable critique without removing educators from the loop. Learners can rehearse critique between live sessions and translate feedback into revisions, while educators can inspect evidence traces such as critique participation, persona selections, exported records, and action-plan generation to see where support may be needed.

This submission does not claim measured learning gains or validated automated assessment. It contributes a working interactive system for artefact-aware AI critique suitable for live demonstration. Future work should evaluate effects on critique quality, reflection, revision behavior, peer feedback, and educator decision-making.

\section*{Acknowledgements}

Critsly was developed with support from the AIEDx grant awarded by IAIED (UEN 53487769M).

\end{document}